\documentclass[twocolumn,amsmath,amssymb]{revtex4}

\def\gsim{ \lower .75ex \hbox{$\sim$} \llap{\raise .27ex \hbox{$>$}} }
\def\lsim{ \lower .75ex \hbox{$\sim$} \llap{\raise .27ex \hbox{$<$}} }

\usepackage{graphicx}
\usepackage{dcolumn}
\usepackage{bm}
\input epsf

\begin{document}

\title{Inflation versus Cyclic Predictions for Spectral Tilt }

\author{ Justin Khoury$^{1}$, Paul J. Steinhardt$^2$ and Neil Turok$^{3}$}

\affiliation{$^1$ ISCAP, Columbia University, New York, NY 10027, USA \\
$^2$ Joseph Henry Laboratories, Princeton University, Princeton, NJ 08544, USA \\
$^3$ DAMTP, Centre for Mathematical Sciences, Wilberforce Road, Cambridge, CB3 0WA, UK}

\begin{abstract}
We present a  nearly model-independent estimate that yields the predictions of  a class of simple inflationary and ekpyrotic/cyclic models for the spectral tilt of the primordial density inhomogeneities that enables us to compare the two scenarios.  Remarkably, we find that the two produce  an identical result, $n_s \approx 0.95$.  For inflation, the same estimate predicts a ratio of tensor to scalar contributions to the low $l$ multipoles of the microwave background anisotropy of $T/S \approx 20$\%; the tensor contribution is negligible for ekpyrotic/cyclic models, as shown in earlier papers.
\end{abstract}

\maketitle


The recent measurement of the cosmic microwave background (CMB) anisotropy by WMAP~\cite{wmap} is consistent with a primordial power spectrum of density fluctuations that is scale-invariant, gaussian and adiabatic. These characteristics coincide with the predictions of the simplest inflationary scenarios~\cite{Guth}. 

In this paper, we show that these are also predicted by the simplest ekpyrotic/cyclic scenarios~\cite{orig,seiberg,ST1}. We compare the density fluctuation spectra obtained in inflationary~\cite{Guth} and ekpyrotic/cyclic models by computing their predictions for an important, well-motivated class of simple models. We find surprisingly similar predictions for the spectral index of the scalar density fluctuations~\cite{g4}. Both predict a red spectrum with index $n_s \approx 0.95$.  For inflation, the same argument predicts a ratio of tensor to scalar contributions to the low $l$ multipoles of roughly 20 percent. Our results for inflation are not new; the particular form of the argument presented here is a variant of the discussion in Ref.~\cite{snow} 
and  by V. Mukhanov~\cite{Mukh1} and 
gives a similar result to other estimates. 
But our result for the ekpyrotic/cyclic models and the 
similarity to the inflationary prediction is both new and unexpected.

For the ekpyrotic/cyclic models, scale-invariant
 fluctuations 
 are generated during a period of slow contraction. 
The notion is that these
imprint themselves as temperature 
fluctuations in the current expanding phase~\cite{ekperts}. 
The validity of this idea has been debated~\cite{ekperts,durrer,brand}, 
with different answers obtained depending on assumptions about 
the precise matching conditions at the bounce. Here we use the recent 
results of Tolley {\it et al.}~\cite{tolley}, which 
treat  the bounce as a collision
of branes in five dimensions, derive  a unique matching 
condition, and find a  scale-invariant spectrum  of temperature 
fluctuations after the bounce.

Both inflation and the ekpyrotic/cyclic models rely on the equation of state parameter $w$ having a specific qualitative behavior throughout the period when fluctuations are generated, including the interval when fluctuations with wavelengths within the present horizon radius were produced (corresponding to the last ${\cal N} \approx 60$ e-folds in wavelength). For inflation, the condition on $w$ is that $1+w \ll 1$ and for ekpyrotic/cyclic models it is $w \gg1$~\cite{ekperts,gratton}. Correspondingly, the Hubble constant $H$ is nearly constant during inflation and the four dimensional scale factor $a$ is nearly constant during ekpyrosis. Since these conditions must be maintained for the duration of an epoch spanning many more than ${\cal N}$ e-folds, the simplest possibility is to suppose that $w$ (and correspondingly $H$ or $a$) change slowly and monotonically during that last ${\cal N}$ e-folds. More precisely, we take ``simplest'' to mean that (i) $d w/d{\cal N}$ is small, and $d^2w/d{\cal N}^2$ or $(d w/d{\cal N})^2$  negligible and (ii) in order for inflation (ekpyrosis) to end, $H$ during inflation (or $a$ during ekpyrosis) decays by a factor of order unity over the last ${\cal N}$ e-folds. Tilts or spectral features that differ from those presented here can only be produced by introducing by hand unnecessary rapid variations in $w$ -- unnecessary in the sense that they are not required for either model to give a successful account of the standard cosmology.

Note that our condition on the time-variation of $w$ does not refer directly to any particular inflaton or cyclic scalar field  potential. In fact, it does not assume that either scenario is driven by a scalar field at all. But, one might ask: how does our condition on the equation of state translate into a condition on an inflaton potential? The answer is simple: it means that the potential  is characterized by a single dimensionful scale, typically $H_I$, the Hubble parameter during inflation. For example, for many models the effective potential is well characterized as $M^4 f(\phi/M)$, where $\phi$ is the inflaton field, $H_I \approx M^2/M_{Pl}$ where $M_{Pl}$ is the Planck mass, and $f(x)$ is a smooth function which, when expanded in $\phi/M$, has all dimensionless parameters of the same order~\cite{wang}. In these cases, to produce inflationary models in which there are rapid changes in the equation of state in the last ${\cal N}$-folds, sharp features have to be introduced in the inflaton potential: bumps, wiggles, steep waterfalls, {\it etc.} But recall that the inflaton field is rolling very slowly throughout inflation, including the last ${\cal N}$ e-folds. Typically, $\phi$ rolls a short distance, $\Delta \phi \ll M$, during the last ${\cal N}$ e-folds. Hence, any sharp features must take place over a range $\delta \phi \ll \Delta \phi \ll M$, or equivalently, by introducing new fields or new mass scales much greater than $M$ in the inflaton potential. For the purposes of comparing the inflationary and ekpyrotic/cyclic predictions, it makes most sense to consider the class with fewest parameters and simplest uniform behavior of the equation of state, a class which is also well-motivated in both models.

Recently, Gratton {\it et al.}~\cite{gratton} analyzed the conditions on the equation of state $w$ required in order for quantum fluctuations in a single scalar field to produce nearly scale-invariant density perturbations, including models which (in the four dimensional effective description) bounce from a contracting to an expanding phase. Their analysis showed that there are only two cases which avoid extreme fine-tuning of initial conditions and/or the effective potential: $w\approx -1$ (inflation) and $w \gg 1$ (the ekpyrotic/cyclic scenario). 

Following Gratton {\it et al.}~\cite{gratton}, we shall discuss the production of long wavelength perturbations in the gauge invariant Newtonian potential $\Phi$, which completely characterizes the density perturbation. Defining $u\equiv a\Phi/\phi'$ (henceforth, primes denote differentiation with respect to conformal time $\tau$), then a Fourier mode of $u$ with wavenumber $k$,  $u_k$, obeys the differential equation
\begin{equation}
u_k'' + \left(k^2 - \frac{\beta\left(\tau\right)}{\tau^2} \right) u_k = 0\,, 
\label{u2}
\end{equation}
with
\begin{eqnarray}
\nonumber
 \beta\left(\tau\right) \equiv \tau^2 H^2 a^2 \left\{\bar{\epsilon} -
\frac{(1-\bar{\epsilon}^2)}{2}\left(\frac{d\ln\bar{\epsilon}}{d{\cal N}}\right)\right. \\\left.\qquad +  \frac{(1-\bar{\epsilon}^2)}{4}\left(\frac{d\ln\bar{\epsilon}}{d{\cal N}}\right)^2 - \frac{(1-\bar{\epsilon})^2}{2}\frac{d^2\ln\bar{\epsilon}} {d{\cal N}^2}\right\}\,,
\label{m}
\end{eqnarray}
where $H=a'/a^2$ is the Hubble parameter, and where $\bar{\epsilon}$ is related to the equation of state parameter $w$ by
\begin{equation}
\bar{\epsilon} \equiv \frac{3}{2}(1+w)\,.
\label{weps}
\end{equation}
We have introduced the dimensionless time variable ${\cal N}$, defined by
\begin{equation}
{\cal N} \equiv \ln\left(\frac{a_{end}H_{end}}{aH}\right) \,,
\label{N}
\end{equation}
where the subscript ``end'' denotes that the quantity is to be evaluated at the end of the inflationary expansion phase
or ekpyrotic contraction phase (corresponding to $w \gg 1$). 
Note that ${\cal N}$ measures the number of e-folds of modes which  exit the horizon before the end of the inflationary or ekpyrotic phase. (N.B.  $d{\cal N} = (\bar{\epsilon}-1) d N$ where $N= {\rm ln} \, a$ in Ref.~\cite{gratton}.) Indeed, defining as usual the moment of horizon-crossing as $k_{\cal N}=aH$ for a given Fourier mode with comoving wavenumber $k_{\cal N}$, then
\begin{equation}
{\cal N} = \ln\left(\frac{k_{end}}{k_{\cal N}}\right) \,,
\end{equation}
where $k_{end}$ is the last mode to be generated.

For nearly constant $w$ (or constant $\bar{\epsilon}$), the unperturbed equations of motion have the approximate solution
\begin{equation}
a(\tau) \sim (-\tau)^{1/(\bar{\epsilon}-1)}\,,\qquad H = \frac{1}{(\bar{\epsilon}-1)a\tau}\,.
\label{sol}
\end{equation}
Substituting the second of these expressions in $\beta$, we find
\begin{equation}
\beta(\tau) \approx \frac{1}{(1-\bar{\epsilon})^2}\left\{\bar{\epsilon} -
\frac{(1-\bar{\epsilon}^2)}{2}\left(\frac{d\ln\bar{\epsilon}}{d{\cal N}}\right) 
\right\}\,,
\end{equation}
where we have assumed that the higher-order derivative terms $d^2\ln\bar{\epsilon}/d{\cal N}^2$ and $(d\ln\bar{\epsilon}/d{\cal N})^2$ are much smaller than $d\ln\bar{\epsilon}/d{\cal N}$.

With the approximation that $\beta$ is nearly constant for all modes of interest, Eq.~(\ref{u2}) can be solved analytically, and the resulting deviation from scale invariance is simply given by the master equation
\begin{equation}
n_s - 1 \approx -2\beta \approx  -\frac{2}{(1-\bar{\epsilon})^2}\left\{\bar{\epsilon} -
\frac{(1-\bar{\epsilon}^2)}{2}\left(\frac{d\ln\bar{\epsilon}}{d{\cal N}}\right) \right\}\,.
\label{ns}
\end{equation}

\noindent{\it Inflation.} Inflation is characterized by a period of superluminal expansion during which $w\approx -1$; that is, $\bar{\epsilon}\ll 1$. In this case, Eq.~(\ref{ns}) reduces to 
\begin{equation}
n_s - 1 \approx -2\bar{\epsilon} + \frac{d\ln\bar{\epsilon}}{d{\cal N}}\,,
\label{nsinf}
\end{equation}
as derived by  Wang {\it et al.}~\cite{wang}.

The next step consists in rewriting the above in terms of ${\cal N}$ only. For this purpose, we need a relation between $\bar{\epsilon}$ and ${\cal N}$. During inflation, the Hubble parameter is nearly constant, but the ``end'' means that $H$ begins to change significantly. So, if we are considering the last ${\cal N}$ e-folds, then, using Eqs.~(\ref{sol}) and the definition of ${\cal N}$ (see Eq.~(\ref{N})), it must be that $H$ decays by a factor of order unity  over those ${\cal N}$ e-folds or
\begin{equation}
\frac{H_{end}}{H}=\left(\frac{a}{a_{end}}\right)^{\bar{\epsilon}}\approx e^{-\bar{\epsilon}{\cal N}} \approx e^{-1}\,,
\label{eq1}
\end{equation}
or
\begin{equation}
\bar{\epsilon}\approx \frac{1}{{\cal N}}\,.
\label{eps1}
\end{equation}

Assuming that this relation holds approximately for all relevant modes, we may substitute in Eq.~(\ref{nsinf}) and obtain
\begin{equation}
(n_s -1)_{inf}\approx -\frac{2}{{\cal N}} - \frac{1}{{\cal N}} = -\frac{3}{{\cal N}}\,.
\label{nsinffinal}
\end{equation}
Note that, in this approximation, the two terms on the right hand side of Eq.~(\ref{nsinffinal}) are both of order $1/{\cal N}$. Figuring that our approximation is good to order $1/{\cal N}$ or a few percent, the result is in agreement with the tilt predicted by simple inflationary models~\cite{phi4}. 

To obtain a numerical estimate of $n_s$, we may derive an approximate value for ${\cal N}$ from the observational constraint that the amplitude of the density perturbations, $\delta\rho/\rho$, be of order $10^{-5}$. In the simplest inflationary models, $\delta\rho/\rho$ is given by~\cite{g4}
\begin{equation}
\frac{\delta\rho}{\rho} \approx \left(\frac{T_r}{M_{Pl}}\right)^2\bar{\epsilon}^{-1/2}\approx 
\left(\frac{T_r}{M_{Pl}}\right)^2 {\cal N}^{1/2}\sim 10^{-5}\,,
\label{coninf1}
\end{equation}
where $T_r$ is the reheat temperature. On the scale of the observable universe today, ${\cal N}$ has the value (see Eq.~(\ref{N}))
\begin{equation}
{\cal N} = \ln\left(\frac{a_{end}H_{end}}{a_0H_0}\right) \approx \ln\left(\frac{T_r}{T_0}\right)\,,
\label{coninf2}
\end{equation}
where $a_0$, $H_0$ and $T_0$ are respectively the current values of the scale factor, Hubble parameter and (photon) temperature. For simplicity, we have 
assumed that $H_{end}\sim T_r^2/M_{Pl}$. 
Combining Eqs.~(\ref{coninf1}) and~(\ref{coninf2}), we obtain the constraint
\begin{equation}
e^{\cal N}{\cal N}^{1/4}\approx 10^{-5/2}\frac{M_{Pl}}{T_0}\,,
\label{Ninf}
\end{equation}
which implies ${\cal N}\approx 60$. It follows that $T_r\sim 10^{16}$~GeV.

If we substitute ${\cal N}\approx 60$ in Eq.~(\ref{nsinffinal}), we obtain $n_s \approx 0.95$, within a percent or two of what is found for the simplest slow-roll and chaotic potentials~\cite{Cope,hodges}.

The prediction for the ratio of tensor to scalar contributions to the quadrupole of the CMB for a  model with 70\% dark energy and 30\% matter is, then,~\cite{Cope,hodges,knox}
\begin{equation}
T/S  \approx  13.8\bar{\epsilon} \approx \frac{13.8}{{\cal N}} \approx 23 \%\,,
\label{rat}
\end{equation}
which is very pleasing because it is in the range which is potentially detectable in the fluctuation spectrum and/or the CMB polarization in the near future~\cite{clt}. (The WMAP collaboration~\cite{peiris} uses a different convention for $T/S$, defining $(T/S)_{WMAP}$ as the ratio of tensor to scalar amplitude of the primordial spectrum. The conversion factor to our $T/S$ is $(T/S)_{WMAP}\approx 1.16\,(T/S)$.)

It is sometimes said that it is easy to construct models where $T/S$ is very small, less than 1\%, say. The argument is that the amplitude of tensor fluctuations is proportional to $H^2$, and a modest decrease in the energy scale for inflation reduces the tensor amplitude significantly. However, one must also consider Eq.~(\ref{rat}) combined with Eq.~(\ref{eps1}). From Eq.~(\ref{rat}), making $T/S$ less than 1\%, for instance, requires $\bar{\epsilon} < 10^{-3}$, which implies ${\cal N} > 1000$. Since we are interested in $T/S$ at ${\cal N} \approx 60$, however, the only way to accommodate such a small $\bar{\epsilon}$ at ${\cal N} \approx 60$ is to have $\bar{\epsilon}$ make a rapid change at some point between ${\cal N} \approx 60$ and the end of inflation. This is precisely what is done in models which yield a small $T/S$ ratio. (Restated in terms of the inflation potential $V(\phi)$, in order to have $T/S\approx 13.8\bar{\epsilon}\approx 28\;(d\ln V/d\phi)^2 \ll 1\%$, it must be that $d\ln V/d\phi \ll 0.02$, which is too small if inflation is to end in 60 e-folds unless one introduces a very rapid change in the slope during the last 60 e-folds.)

\noindent{\it Ekpyrotic/cyclic models.} The ekpyrotic phase is characterized by a period of slow contraction with $w\gg 1$; that is, $\bar{\epsilon}\gg 1$. In this case, Eq.~(\ref{ns}) reduces to~\cite{gratton}
\begin{equation}
n_s -1 \approx -\frac{2}{\bar{\epsilon}} - 
\frac{d\ln\bar{\epsilon}}{d{\cal N}}\,.
\label{nsek}
\end{equation}
Notice that this relation can be transformed into the expression for inflation, Eq.~(\ref{nsinf}), by replacing $\bar{\epsilon} \rightarrow 1/\bar{\epsilon}$.  Note further that, for all cosmologies, the scale factor is $a \propto t^{1/\bar{\epsilon}} \propto H^{-1/\bar{\epsilon}}$, where $t$ is proper time. Hence,  inflation ($\bar{\epsilon}\ll 1$) has $a$ rapidly varying and $H$ nearly constant, whereas the ekpyrotic/cyclic model ($\bar{\epsilon}\gg 1$) has $H$ varying and $a$ nearly constant. This suggests an interesting duality between the inflationary and ekpyrotic/cyclic models that reflects itself in the final results.

If the scale factor $a(\tau)$ is nearly constant during the ekpyrotic (contraction) phase, then the phase ends when $a(\tau)$ begins to change significantly. In particular, the condition that the scale factor $a(\tau)$ decays by a factor of order unity during the last ${\cal N}$ e-folds reads 
\begin{equation}
\frac{a_{end}}{a} = \left(\frac{aH}{a_{end}H_{end}}\right)^{1/(\bar{\epsilon}-1)}\approx e^{-{\cal N}/\bar{\epsilon}}\approx e^{-1}\,
\label{decayek}
\end{equation}
(the analogue of Eq.~(\ref{eq1}) for inflation), which implies
\begin{equation}
\bar{\epsilon}\approx {\cal N}\,
\label{eps2}
\end{equation}
(to be compared with Eq.~(\ref{eps1}) for inflation). Substituting this expression into Eq.~(\ref{nsek}), one obtains
\begin{equation}
(n_s - 1)_{ek} \approx -\frac{2}{{\cal N}} - \frac{1}{{\cal N}} = - \frac{3}{{\cal N}}\,.
\label{nsekfinal}
\end{equation}
This is the key relation for the ekpyrotic/cyclic models. 

In the inflationary case, we estimated ${\cal N}$ by using the constraint on the amplitude of density perturbations, $\delta\rho/\rho\sim 10^{-5}$. 
For ekpyrotic and cyclic models, this constraint
involves more parameters and is therefore not 
sufficient by itself 
to fix ${\cal N}$~\cite{tolley,design}. To estimate ${\cal N}$, 
we rewrite Eq.~(\ref{N}) as
%
\begin{equation}
{\cal N} \approx \ln\left(\frac{T_r}{T_0}\right) + \ln\left(\frac{a_{end}H_{end}}{a_rH_r}\right)\,,
\label{rewriteN}
\end{equation}
where the subscript $r$ denotes the onset of the radiation-dominated phase. 
In inflation, we have $a_{end}\approx a_r$ and $H_{end}\approx H_r$. 
In the ekpyrotic/cyclic model, however,
the end of ekpyrosis occurs during the contracting phase whereas the onset of radiation-domination is during the expanding phase.
To  estimate the ratio $a_{end}H_{end}/a_r H_r$, we note
that, from approximately the end of ekpyrosis, through the bounce,
and up to the onset of radiation-domination, the universe is dominated by 
scalar field kinetic energy, {\it i.e.}, $w\approx 1$ \cite{seiberg,ST1}. 
From Eqs.~(\ref{weps}) and~(\ref{sol}), we find $a\approx (-\tau)^{1/2}\sim H^{-1/3}$, and therefore
\begin{equation}
\frac{a_{end}H_{end}}{a_rH_r} \approx \left(\frac{H_{end}M_{Pl}}{T_r^2}\right)^{2/3}\,.
\end{equation}
Substituting in Eq.~(\ref{rewriteN}), we find
\begin{equation}
e^{\cal N} = \left(\frac{H_{end}^2}{T_rM_{Pl}}\right)^{1/3}\frac{M_{Pl}}{T_0}\,,
\label{Nek}
\end{equation}
which is the analogue of Eq.~(\ref{Ninf}). 

The constraints on $H_{end}$ and $T_r$ in cyclic models are analyzed
in Ref.~\cite{design} 
and the  range 
of allowed values is presented.
Central values  are $T_r\approx 10^5$ GeV and $H_{end}\approx 10^5$ GeV, which, from Eq.~(\ref{Nek}), implies ${\cal N}\approx 60$. 
(By pushing parameters, ${\cal N}$ can be made to vary 20\% or so one
way or the other.)
Substituting ${\cal N}=60$ in the expression for the tilt 
gives $n_s \approx 0.95$, the same estimate obtained for inflation.

\noindent{\it Conclusions.} Remarkably, our estimates for the typical tilt in the inflationary and ekpyrotic/cyclic models are virtually identical. Both models predict a red spectrum, with spectral slope
\begin{equation}
n_s-1\approx - \frac{3}{{\cal N}}\,.
\end{equation}
Furthermore, when adding observational constraints such as the COBE constraint that the amplitude of density fluctuations be of order $10^{-5}$, both models yield ${\cal N} \approx 60$. This results in an identical prediction for the spectral tilt of $n_s\approx 0.95$. Furthermore, in both models, the time-variation of the equation of state contributes a correction of ${\cal O}(1)$ that reddens the spectrum. We have seen that this occurs because there is fascinating duality ($\bar{\epsilon} \rightarrow 1/\bar{\epsilon}$) between inflationary and ekpyrotic/cyclic conditions. This result was neither planned nor anticipated and suggests a deep connection between the expanding inflationary phase and the contracting ekpyrotic/cyclic phase. The key difference is that inflation also predicts a nearly scale-invariant spectrum of gravitational waves with a detectable amplitude. The predicted ratio of tensor to scalar CMB multipole moments at low $l$ is  $T/S \approx 20$\%. The tensor spectrum from cyclic models is strongly blue and exponentially small on cosmic scales~\cite{orig,boyle}.

We thank V. Mukhanov for useful discussions. This work was supported in part by US Department of Energy grants DE-FG02-91ER40671 (PJS) and DE-FG02-92ER40699 (JK), the Columbia University Academic Quality Fund (JK), the Ohrstrom Foundation (JK), and PPARC, UK (NT).


\begin{thebibliography}{9999}
\bibitem{phi4} In fact it is precisely the result predicted for a $\phi^4$ inflationary model.
\end{thebibliography}
\end{document}